\pdfoutput=1
\documentclass[a4paper,10pt,notitlepage,reqno]{article}
\usepackage[utf8]{inputenc}
\usepackage[english]{babel}
\usepackage{amssymb,amsthm,bm}
\usepackage{mathtools}
\usepackage{mathrsfs}
\usepackage{enumitem}
\usepackage{authblk}
\usepackage{xcolor}
\usepackage{tikz}
\usepackage{stackrel}
\usepackage{geometry}
\usepackage{natbib}
\usepackage{graphicx}
\usepackage{subcaption}
\usepackage{hyperref}
\usepackage{moreverb,url}
\usepackage{cancel}

\def\1{{\rm l}\hskip -0.21truecm 1}
\usepackage{xcolor}
\definecolor{verd}{HTML}{94CCAB}

\usepackage{sectsty}
\sectionfont{\Large}
\subsectionfont{\large}

\title{
Bayesian joint modeling of longitudinal patient-reported outcomes and survival: an application to chronic obstructive pulmonary disease}

\author[1,2,*]{Cristina Gal\'an-Arcicollar}
\author[3]{Danilo Alvares}
\author[2]{Josu Najera-Zuloaga}
\author[4]{Dae-Jin Lee}

\affil[1]{Applied Statistics, Basque Center for Applied Mathematics BCAM, Bizkaia, Spain}
\affil[2]{{Department of Mathematics, University of the Basque Country UPV/EHU, Bizkaia, Spain}}
\affil[3]{{MRC Biostatistics Unit, University of Cambridge, Cambridge, UK}}
\affil[4]{{School of Science and Technology, IE University, Madrid, Spain}}

\affil[*]{\textit{Corresponding author:} cristina.galanarcicollar@gmail.com}

\date{}

\begin{document}
\maketitle

\begin{abstract}

Questionnaire-based patient-reported outcomes (PROs) are discrete, bounded and overdispersed, yet joint models relating them to survival may ignore these features or estimate both processes sequentially. We propose a Bayesian joint model combining a beta-binomial mixed-effects submodel with a Weibull proportional hazards submodel, linked through the subject-specific response probability. Simulations show that simultaneous estimation reduces bias in the longitudinal slope and yields practically unbiased association estimates, unlike two-stage estimation. In a cohort of 543 patients with chronic obstructive pulmonary disease, the model identified associations for all eight SF-36 dimensions and for two of three SGRQ dimensions, including several associations not detected by the two-stage approach, and provided dynamic survival predictions.

 \end{abstract}

\vspace{1em}


\section{Introduction}

A patient-reported outcome (PRO) is any report on the status of a patient’s health condition, quality of life, or functional status associated with healthcare or treatment that comes directly from the patient, without interpretation of the patient’s response by a clinician or anyone else \citep{Services2006}. PROs are increasingly being used as primary outcome measures in both observational and experimental studies due to their ability to provide valuable insights into patients' health. These data help to improve patient care and contribute to the development of more patient-centered healthcare systems by measuring outcomes that are difficult or impossible to assess through clinical measurements, such as pain, quality of life, or satisfaction with care. Their use is strongly recommended in combination with clinical indicators to provide a more comprehensive patient assessment, especially in chronic illnesses \citep{Speight2010}.

Among the different types of PROs, health-related quality of life (HRQoL) is particularly relevant, as it captures the impact of health status on an individual’s physical, psychological, and social functioning. Clinical studies typically collect repeated HRQoL measurements to monitor longitudinal changes, and substantial evidence indicates that these scores are also valuable prognostic indicators for clinical outcomes such as mortality \citep{Kluetz2016}. {Consequently, HRQoL measurements are increasingly collected alongside time-to-event outcomes, such as death or disease progression.} In this context, regression models are essential for assessing the relationship between changes in HRQoL and patient survival.

When research focuses on the relationship between repeated longitudinal measurements and time-to-event data, joint models provide an effective statistical framework. The core idea is to link a survival submodel with a {mixed-effects} submodel for repeated measurements, enabling a more integrated analysis. Usually, these two submodels are connected via shared random effects or functions of the underlying longitudinal trajectory \citep{Rizopoulos2012}. Compared with separate analyses, joint models use the information from both processes simultaneously and can reduce bias and improve efficiency when the longitudinal and event processes are associated. {They may also be preferable to standard Cox models that directly incorporate the observed longitudinal marker as a time-dependent covariate, because joint models account for measurement error and for the fact that the marker is observed only intermittently} \citep{Arisido2019}. Moreover, joint models enable clinically relevant applications such as personalized predictive modeling, allowing clinicians to update an individual patient’s risk as new longitudinal information becomes available, making them particularly useful for developing effective and individualized screening and monitoring strategies \citep{Andrinopoulou2021}.

{In this work, we focus on questionnaire-based PRO data, with scores typically obtained as sums or weighted combinations of responses to multiple items. Many of these scores are therefore discrete and bounded random variables and may exhibit additional variability beyond that allowed by a binomial distribution.} In this context, the beta-binomial distribution has been proposed as a suitable model for these types of data \citep{Arostegui2007}, as it accounts for the overdispersion observed in PROs compared with more classical discrete and bounded alternatives, such as the binomial distribution. {Within the joint modeling framework, several methodological developments have been proposed over recent decades to accommodate non-Gaussian longitudinal outcomes and alternative survival specifications \citep{Wang2026}. A relevant contribution for discrete and bounded longitudinal outcomes was provided by \citet{VandenHout2016}, who combined binomial and beta-binomial mixed-effects models with parametric survival models and performed simultaneous estimation by marginal likelihood. Their shared-parameter formulation links the two processes through the subject-specific random effects. Although this provides a valid dependence structure, the resulting association is expressed in terms of latent subject-specific effects rather than directly through the underlying response probability, which may be less natural to interpret in the context of PROs. Beta-binomial longitudinal submodels are also available in general-purpose Bayesian joint modeling software, such as the \texttt{JMbayes2} R package \citep{JMbayes2}. However, to our knowledge, the extent to which simultaneous estimation corrects the biases of the two-stage strategies used in applied PRO research has not been systematically evaluated.}

{Despite these developments, the clinical application of joint models for assessing the relationship between longitudinal HRQoL and mortality remains limited \citep{Yonemoto2026}. For example, longitudinal changes in HRQoL have been related to subsequent mortality without jointly modeling the longitudinal and complete time-to-event processes \citep{Orive2023}, whereas other analyses have modeled vital status without incorporating the complete time-to-event information \citep{Esteban2022}. Other contributions have transformed HRQoL scores to accommodate Gaussian longitudinal submodels \citep{Ibrahim2010, Li2013, Daza2024} or linked the longitudinal and survival processes only through shared random effects \citep{Li2010, Choi2015, Hatfield2011}. Although these association structures account for dependence between the two processes, their interpretation in terms of the underlying HRQoL trajectory is less direct. A recent two-stage approach incorporated the beta-binomial distribution and linked the estimated response probability to survival, providing a more direct response-scale interpretation \citep{Galan-Arcicollar2024}. However, its longitudinal submodel is estimated without using survival information, and the resulting estimated trajectories are subsequently treated as known in the survival model. Consequently, mortality-related dropout cannot inform estimation of the longitudinal trajectory, and longitudinal estimation uncertainty is not fully propagated into the association parameter. These features may produce bias in the longitudinal parameters and in the estimated association between HRQoL and mortality \citep{Wu2012, Ye2008}.}

{To address these limitations, this paper formulates a fully integrated Bayesian joint modeling approach for longitudinal PRO and survival data. The beta-binomial distribution accommodates the discrete, bounded, and overdispersed characteristics of the longitudinal outcomes, while the survival and longitudinal submodels are linked directly through the underlying response probability trajectory. This linkage provides a natural interpretation of the association between the patient’s underlying PRO level and mortality risk. Moreover, simultaneous estimation allows survival information to contribute to the estimation of longitudinal trajectories and propagates longitudinal estimation uncertainty into the survival component. A simulation study evaluates whether this formulation reduces bias in the longitudinal parameters relative to independent longitudinal modeling and in the association parameter relative to two-stage estimation.} Although joint models can be estimated under both frequentist and Bayesian paradigms, we adopt a Bayesian framework for two main reasons. First, the non-Gaussian nature of PRO data, the hierarchical structure of the beta-binomial longitudinal submodel, and the nonlinear association between the longitudinal process and the hazard function via the inverse logit of the linear predictor lead to a highly complex likelihood, whose marginalization over the random effects is more naturally handled through MCMC sampling than through numerical integration. Second, although dynamic predictions can also be obtained from frequentist joint models \citep{Proust-Lima2016}, the Bayesian framework propagates parameter uncertainty more straightforwardly because it naturally yields posterior samples of all model parameters, directly enabling subject-specific dynamic predictions.

{The proposed framework is illustrated using data from a longitudinal observational study conducted by researchers at the Respiratory Service of Galdakao Hospital in Biscay, Spain, which aimed to track the health status and progression of patients with chronic obstructive pulmonary disease (COPD) \citep{Esteban2020}.} COPD is a complex, heterogeneous, and multisystem disease that extends beyond chronic airflow limitation, causing both physiological and psychosocial discomfort \citep{Vanfleteren2016}.  {The study collected repeated HRQoL measurements together with mortality information, providing an appropriate setting in which to investigate the relationship between patients' HRQoL trajectories and survival.} Indeed, HRQoL assessment is already considered part of standard care in COPD treatment \citep{Willke2004}. {The same cohort was previously analyzed using longitudinal methods and the two-stage approach \citep{Esteban2020, Galan-Arcicollar2024}. In the present work, it serves as an illustrative application of the simultaneous Bayesian framework, allowing the resulting longitudinal and survival inferences to be compared with those obtained when the two processes are not estimated simultaneously.}

Relative to the two-stage approach of \citet{Galan-Arcicollar2024}, applied to the same cohort, this work makes three contributions: (i) a simultaneous Bayesian formulation that links the survival submodel directly to the beta-binomial response probability and accounts for delayed entry; (ii) a simulation study quantifying the bias of two-stage estimation in the longitudinal slope and the association parameter under mortality-related dropout; and (iii) subject-specific dynamic survival predictions that propagate parameter uncertainty.

The paper is structured as follows. {Section \ref{Section2} introduces the illustrative COPD study and provides a description of the dataset.} Section \ref{Section3} presents the joint model, incorporating the beta-binomial distribution as the longitudinal submodel. {Section \ref{Section4} presents the simultaneous Bayesian inference procedure; the previously proposed two-stage approach \citep{Galan-Arcicollar2024} is summarized as the benchmark used for comparison.} Section \ref{Section5} applies the proposed method to the COPD dataset and illustrates dynamic predictions based on the model. {Section \ref{Section6} presents a simulation study that evaluates longitudinal parameter estimation relative to the independent first-stage fit and association parameter estimation relative to the complete two-stage procedure.} Finally, Section \ref{Section7} offers conclusions and discussion.

\section{COPD study and data}\label{Section2}

{COPD is a heterogeneous lung condition characterized by chronic respiratory symptoms and persistent, often progressive, airflow obstruction due to abnormalities of the airways and/or alveoli \citep{GOLD2023}.} It is considered a major cause of chronic morbidity and mortality worldwide \citep{Pauwels2004}. {Based on available epidemiological studies, its global prevalence has been estimated at 13.1\%, highlighting the substantial worldwide burden of the disease \citep{Blanco2019}.}

Although chronic airflow limitation defines COPD, the disease is recognized as complex, heterogeneous, and multicomponent, causing not only physiological discomfort but also a profound psychosocial impact on individuals. {Standard clinical evaluations, typically focused on physiological and clinical measurements, may therefore fail to capture the full burden of the disease.} {Consequently, no single clinical measure can fully reflect the severity or progression of COPD, and its assessment should be complemented by patient-centered indicators, particularly HRQoL, which is considered an important component of COPD care \citep{Willke2004}.}

{The data analyzed in this work arise from a prospective observational study conducted by the Respiratory Service at Galdakao Hospital in Biscay, Spain \citep{Esteban2020}.} {A total of 543 patients were consecutively enrolled during the first 18 months of the study if they had been diagnosed with COPD for at least six months and had been clinically stable for six weeks before enrollment.} They were followed for a five-year period, during which four clinical examinations and interviews were planned per patient. {The number of recorded assessments varied because some patients died or were lost to follow-up before completing all four planned visits, resulting in an unbalanced longitudinal dataset.} Consequently, the number of measurements per patient ranges from one to four, totaling 1,772 observations. Additionally, the timing of patients’ first measurements varied due to staggered enrollment, and the intervals between measurement times were uneven. Specifically, the second and third measurements were scheduled one year apart, but the fourth measurement was conducted three years after the third.

{The event of interest was patient mortality, with 167 deaths recorded during follow-up.} {The remaining 376 patients were right-censored: 324 were administratively censored at the end of the observation period, whereas 52 were censored because they were lost to follow-up before study completion.}

In the COPD study, health status was assessed using both generic and disease-specific questionnaires: the Short Form-36 Health Survey (SF-36) and the St. George’s Respiratory Questionnaire (SGRQ), respectively. These questionnaires capture various aspects of health, categorized into different dimensions based on the specific information they provide.

The SF-36 is one of the most widely used instruments for assessing HRQoL across diverse populations and conditions \citep{Ware1993}. {It consists of 36 items grouped into eight distinct dimensions.} {Four dimensions primarily assess physical aspects of health: \textit{Physical Functioning} (PF), \textit{Role Physical} (RP), \textit{Bodily Pain} (BP), and \textit{General Health} (GH).} {The other four primarily assess mental and social aspects: \textit{Vitality} (VT), \textit{Social Functioning} (SF), \textit{Role Emotional} (RE), and \textit{Mental Health} (MH).} {For each dimension, responses to the corresponding items are combined according to a dimension-specific scoring procedure.} The resulting raw scores are then transformed into standardized scale scores from 0 to 100, with a higher score indicating better health status.

The SGRQ is a self-completed standardized questionnaire widely used to quantify the impact of chronic obstructive airway diseases on symptoms, functional capacities, and general well-being \citep{Doll2003,Peruzza2003}. This questionnaire contains 50 items divided into two main parts: the first includes a single dimension called \textit{Symptoms} (SYMP), while the second encompasses two dimensions: \textit{Impacts} (IMP) and \textit{Activity} (ACT). {Each of the three dimensions is scored separately on a scale from 0 to 100, with higher scores indicating poorer health status.}

In this study, the standardized 0–100 scores provided by both questionnaires were subsequently rescaled to discrete scales that reflect the original structure of the questionnaires, since the dimension scores arise from sums of binary or ordinal items. This transformation allows the outcomes to be modeled on their natural discrete scale and facilitates interpretation. The rescaling of the SF-36 dimensions follows the approach proposed by \cite{Arostegui2013}, while the SGRQ rescaling is based on the notion that a 4-point change on the 0–100 scale represents a clinically meaningful difference \citep{Jones2005}.

\section{Joint model for longitudinal PRO and survival data}\label{Section3}

Joint models are a powerful tool for incorporating shared information between longitudinal and survival processes. {The underlying idea is to specify a mixed-effects submodel for the repeated measurements and a time-to-event submodel, and to link them through shared latent quantities or functions of the underlying longitudinal trajectory \citep{Wulfsohn1997}.} {When the two processes are associated, estimating them simultaneously can reduce bias in parameter estimates and improve variance estimation compared with separate analyses \citep{Faucett1996}.} However, it is essential to fit suitable models that adequately reflect the characteristics of the longitudinal data.

{In this work, we investigate the association between longitudinal PRO trajectories and survival by linking the two processes through the underlying probability trajectory of the longitudinal outcome.} PROs are often constructed as the sum of responses to multiple questionnaire items, making them discrete and bounded random variables. The literature has shown that PROs usually exhibit extra variability beyond the mean-variance structure of the binomial distribution, a property known as overdispersion \citep{Arostegui2007}. {They may also accumulate values at one or both ends of the scale, resulting in shapes such as U-, J-, or reverse J-shaped distributions.} {The beta-binomial distribution has been proposed as an appropriate model for analyzing discrete and bounded outcomes with overdispersion, particularly in the analysis of PROs \citep{NajeraZuloaga2018}.} {It can be represented hierarchically as a binomial distribution conditional on a probability parameter that follows a beta distribution.} Additional details can be found in the literature \citep{Johnson2005}.

Next, we outline the two submodels that constitute the joint modeling framework employed in this study.

\subsection{Longitudinal submodel: beta-binomial mixed-effects model}\label{Subsection3.1}

Let $\bm{y_{i}}=(y_{i1},\ldots,y_{in_{i}})^\prime$ be the vector of $n_{i}$ longitudinal measurements for subject $i$, taken at times $(t_{i1},\ldots,t_{in_{i}})^\prime$, such that $y_{ij}=y_i(t_{ij})$, for $i=1,\ldots,n$ and $j=1,\ldots,n_i$. For each subject $i$, we assume that, conditional on subject-specific random effects, the longitudinal responses are independent and follow a beta-binomial distribution:
\begin{equation*}
y_{ij} \mid \bm{b_{i}} \sim BB(m,p_{ij},\phi) \quad \text{with} \quad \bm{b_{i}} \sim \mathcal{N}(0,\Sigma), \quad j=1,\ldots,n_i \quad \text{and} \quad i=1,\ldots,n,
\end{equation*}
{where $\bm{b}_{i}=(b_{i0},b_{i1})^\prime$ is the subject-specific random-effects vector and $\Sigma$ is its variance-covariance matrix, which induces dependence among the repeated measurements from the same subject.} {The beta-binomial parameter $m$ is the maximum possible score, so that $y_{ij}\in\{0,\ldots,m\}$, whereas $p_{ij}=p_i(t_{ij})$ is the probability parameter for subject $i$ at time $t_{ij}$.} {Under this parameterization, the conditional mean is

$$
E(y_{ij}\mid\bm{b}_i)=m p_{ij},
$$

and therefore \(p_{ij}\) represents the subject-specific expected score expressed as a proportion of the maximum possible score.} {In the PRO context, $m$ is fixed by the questionnaire dimension being analyzed, whereas $p_{ij}$ characterizes the underlying health-status trajectory on a common scale between 0 and 1.} {Finally, $\phi>0$ is the dispersion parameter, with larger values representing greater extra-binomial variability.}

{The beta-binomial mixed-effects model has been developed in detail elsewhere \citep{NajeraZuloaga2019}, where the probability parameter is related to a linear predictor through a logit link function.} {In this work, we adopt this specification for the longitudinal submodel and connect $p_{ij}$ with the linear predictor as follows:}
\begin{equation}
\label{eq:linearBBMM}
\log\left(\dfrac{p_{ij}}{1-p_{ij}}\right) = (\beta_{0} + b_{i0}) + (\beta_{1} + b_{i1})t_{ij},
\end{equation}
{where $\beta_{0}$ and $\beta_{1}$ are the population-level fixed intercept and slope, respectively, whereas $b_{i0}$ and $b_{i1}$ represent the corresponding subject-specific deviations.}

{For the sake of clarity and to isolate the estimation of the longitudinal trend and its association with the survival process, the linear predictor in Equation \eqref{eq:linearBBMM} includes only time as a covariate, although the proposed modeling framework can be straightforwardly extended to incorporate additional baseline covariates, time-varying effects, and nonlinear functions of time. This simplified specification is particularly convenient for our purposes, since it allows us to directly assess whether jointly modeling the longitudinal and survival processes corrects the bias in the estimation of the longitudinal trend that arises from mortality-related dropout, an effect that would be harder to isolate in a covariate-adjusted model.}

{The conditional probability mass function of the longitudinal outcome is}
\begin{equation*}
f(y_{ij} \mid \bm{b_{i}}; \bm{\theta_{y}})=\binom{m}{y_{ij}}\dfrac{\Gamma\left(\dfrac{1}{\phi}\right)}{\Gamma\left(\dfrac{1}{\phi}+m\right)}\dfrac{\Gamma\left(\dfrac{p_{ij}}{\phi}+y_{ij}\right)}{\Gamma\left(\dfrac{p_{ij}}{\phi}\right)}\dfrac{\Gamma\left({\dfrac{1-p_{ij}}{\phi}+m-y_{ij}}\right)}{\Gamma\left(\dfrac{1-p_{ij}}{\phi}\right)},
\end{equation*}
{where $\Gamma(\cdot)$ denotes the gamma function, which satisfies $\Gamma(n)=(n-1)!$ for positive integers $n$.} {The parameters of the longitudinal submodel are collected in $\bm{\theta}_{y}=(\beta_{0},\beta_{1},\phi)^\prime$, whereas the parameters of the random-effects distribution are denoted by $\theta_{b}=\Sigma$.}

\subsection{Survival submodel: proportional hazards model}\label{Subsection3.2}

{Let $(T_{i},\delta_{i})$ be the recorded survival data for the $i$th subject, where $T_i$ represents the observed event or censoring time and $\delta_i$ is the event indicator, taking the value 1 for an event and 0 for censoring, for $i=1,\ldots,n$.}

We define a proportional hazards model for the survival data as follows:
\begin{equation}
\label{eq:survival_submodel}
\lambda_{i}(t \mid \bm{b_{i}};\bm{\theta_{y}},\bm{\theta_{t}})= \lambda_{0}(t)\exp(\alpha~p_{i}(t)) \quad \text{with} \quad \lambda_{0}(t)= \nu t^{\nu-1}\exp(\gamma), \quad i=1,\ldots,n,
\end{equation}
where $\lambda_i$ is the hazard function for the $i$th subject, $\lambda_0(\cdot)$ is the Weibull baseline hazard function defined by parameters $\nu$ and $\gamma$, and $\alpha$ is the association parameter linking the longitudinal and survival processes. The parameters of the survival submodel are collected in $\bm{\theta_{t}}=(\nu,\gamma,\alpha)$. The Weibull baseline hazard is assumed due to its flexibility; its adequacy was assessed using standard residual diagnostics, including Cox–Snell and martingale residuals, which indicated a satisfactory fit to the data.

{In this specification, the hazard at time $t$ depends on $p_i(t)$, obtained by applying the inverse-logit function to the linear predictor defined in Equation~\eqref{eq:linearBBMM}.} {Since $p_i(t)$ represents the subject-specific expected PRO score as a proportion of its maximum possible value, $\exp(\alpha\Delta)$ is the hazard ratio associated with an increase of $\Delta$ in this underlying probability trajectory.} {The model therefore links the longitudinal and survival processes through a directly interpretable feature of the PRO trajectory.}

{Because patients in the COPD study were enrolled at different calendar times, the survival data exhibit delayed entry, which requires accounting for left truncation.} {Let $L_i$ denote the entry time of subject $i$ and $S_i(t\mid\bm{b}_i;\bm{\theta_y},\bm{\theta_t})$ the survival function associated with the hazard in Equation~\eqref{eq:survival_submodel}.} {Conditional on the subject having remained event-free until entry time $L_i$, the likelihood contribution for the survival outcome is formulated as:}

{
\begin{equation}\label{eq::jm_survpart}
\begin{split}
f(T_{i},\delta_{i} \mid  L_{i}, \bm{b_{i}};\bm{\theta_{y}},\bm{\theta_{t}})
&= \lambda_{i}(T_{i})^{\delta_{i}}
\dfrac{S_i(T_{i})}{S_i(L_{i})} \\
&= [\lambda_{0}(T_{i})\exp(\alpha~p_{i}(T_{i}))]^{\delta_{i}}
\dfrac{\exp\left(-\int_{0}^{T_{i}}\lambda_{0}(s)\exp(\alpha~p_{i}(s))ds\right)}
{\exp\left(-\int_{0}^{L_{i}}\lambda_{0}(s)\exp(\alpha~p_{i}(s))ds\right)}.
\end{split}
\end{equation}
}

{The denominator $S_i(L_i)$ conditions the contribution on the subject having remained event-free until entering the study.}

\section{Joint model inference}\label{Section4}

{To estimate the full parameter vector $\bm{\theta}=(\bm{\theta_{y}},\bm{\theta_{t}},\theta_{b})$ together with the shared random effects
$\bm{b}=(\bm{b}_{1}^{\prime},\ldots,\bm{b}_{n}^{\prime})^{\prime}$, the joint modeling framework typically defines a complete likelihood function under the assumption of full conditional independence.} Specifically, it is assumed that, conditional on the random effects, the  longitudinal and time-to-event outcomes are independent, as are the repeated measurements within each individual \citep{Rizopoulos2010}. {Given the collected data $(\bm{y}_{i})_{i=1}^{n}$ and $(T_i,\delta_i,L_i)_{i=1}^{n}$ for $n$ subjects, the complete
likelihood function is:}

{
\begin{equation*}
L(\bm{\theta},\bm{b})=\prod_{i=1}^{n}
\left[f(T_i,\delta_i\mid L_i,\bm{b}_i;
\bm{\theta_y},\bm{\theta_t})
\prod_{j=1}^{n_i}
f(y_{ij}\mid\bm{b}_i;\bm{\theta_y})
f(\bm{b}_i;\theta_b)\right].
\end{equation*}
}

{The marginal likelihood function for the parameter set is then  obtained by integrating out the random effects:}

{
\begin{equation}\label{eq::jm_marginal_likelihood}
L(\bm{\theta})=\prod_{i=1}^{n}
\left[\int f(T_i,\delta_i\mid L_i,\bm{b}_i;
\bm{\theta_y},\bm{\theta_t})
\prod_{j=1}^{n_i}
f(y_{ij}\mid\bm{b}_i;\bm{\theta_y})
f(\bm{b}_i;\theta_b)
\,d\bm{b}_i\right].
\end{equation}
}

{For inference and comparison, we consider two estimation strategies. First, we consider a two-stage strategy (hereafter TSBB) that estimates the longitudinal submodel and subsequently incorporates the fitted subject-specific probability trajectory, evaluated at the relevant event times, as an observed time-dependent covariate in the survival submodel. This strategy provides a natural benchmark because it accounts for the distributional characteristics of PRO data through a beta-binomial longitudinal submodel and links the longitudinal and survival processes through the estimated probability parameter, thereby retaining a direct interpretation of their association \citep{Galan-Arcicollar2024}. We then present the proposed simultaneous Bayesian joint specification  (hereafter JMBB), in which survival information contributes to the estimation of the longitudinal trajectory and uncertainty is propagated jointly across both submodels. As shown in the simulation study in Section~\ref{Section6}, this simultaneous strategy reduces bias in the longitudinal trend estimates relative to independent longitudinal modeling and in the association parameter $\alpha$ relative to two-stage estimation.}

\subsection{A two-stage approach}

{The two-stage strategy considered here follows the approach described in \cite{Galan-Arcicollar2024} and provides a computationally convenient alternative to directly maximizing the marginal likelihood in Equation~\eqref{eq::jm_marginal_likelihood}.}

The first stage involves fitting the longitudinal beta-binomial mixed-effects model to assess the evolution of the longitudinal outcome over time. The beta-binomial distribution does not belong to the exponential family. {Consequently, standard estimation procedures for generalized linear mixed models (GLMMs) \citep{McCulloch2001}, which rely on exponential-family assumptions, cannot be applied directly.} {Instead, parameter estimation is performed using the likelihood-based approach specifically developed for beta-binomial mixed-effects models by \cite{NajeraZuloaga2019}.} This approach allows for the incorporation of random effects while accounting for the extra-binomial variability inherent to the beta-binomial distribution.

{In the second stage, once the longitudinal submodel has been fitted, the estimated subject-specific probability trajectory is treated as an observed time-dependent covariate in a Cox proportional hazards model. At any time $t$, this estimated probability is given by:}
{
\begin{equation*}
\hat{p}_{i}(t)
=
\operatorname{logit}^{-1}
\left((\hat{\beta}_{0}+\hat{b}_{i0})
+(\hat{\beta}_{1}+\hat{b}_{i1})t\right),
\end{equation*}
}
{where $\hat{\beta}_{0}$ and $\hat{\beta}_{1}$ represent the estimated fixed effects, and $\hat{b}_{i0}$ and $\hat{b}_{i1}$ represent the corresponding subject-specific random effects obtained in the first stage.}

{The survival submodel in the second stage is therefore specified as:}
{
\begin{equation*}
\lambda_i(t)
=
\lambda_0(t)\exp\left(\alpha~\hat{p}_i(t)\right).
\end{equation*}
}
{The association parameter $\alpha$ is estimated by partial likelihood maximization using standard Cox proportional hazards regression, leaving the baseline hazard function unspecified \citep{Cox1972}. At each observed event time, $\hat p_i(t)$ is evaluated for every subject in the corresponding risk set. Left truncation is incorporated by defining the observation interval for subject $i$ as $[L_i,T_i]$, so that the subject only contributes to risk sets after entering the study.}

\subsection{Joint specification}

{Although the two-stage strategy provides a computationally practical way to combine beta-binomial longitudinal modeling with survival analysis, its sequential estimation may produce biased estimates \citep{Wu2012}. To address this limitation, we adopt a simultaneous Bayesian joint modeling formulation that retains the beta-binomial distribution for the longitudinal process and estimates the longitudinal and survival parameters jointly.}

Within the Bayesian framework, the objective is to estimate the posterior distributions of the model parameters. These posterior distributions are derived using Bayes’ theorem, which combines the information from the observed data with prior knowledge specified through prior distributions. Such priors may be informed by previous studies, expert knowledge, or reasonable assumptions about the parameters \citep{Gelman2013}.

{For the joint modeling of longitudinal and survival data, the joint posterior distribution of all model parameters $\bm{\theta}=(\bm{\theta_y},\bm{\theta_t},\theta_b)$ and random effects $\bm b$, conditional on the observed data $D_n$, is denoted by $\pi(\bm{\theta},\bm b\mid D_n)$ and satisfies:}

\begin{equation*}
  	\pi(\bm{\theta},\bm{b}\mid D_n) \propto \prod_{i=1}^{n}\left[f(T_{i},\delta_{i}, L_{i} \mid \bm{b_{i}};\bm{\theta_{t}})\prod_{j=1}^{n_{i}}f(y_{ij} \mid \bm{b_{i}};\bm{\theta_{y}})f(\bm{b_{i}};\theta_{b})\right]\pi(\bm{\theta}),
\end{equation*}

{Here, $\pi(\bm{\theta})$ represents the joint prior distribution of the model parameters.}

Since this posterior distribution is typically intractable in closed form, inference is usually performed using Markov chain Monte Carlo (MCMC) techniques. {In this work, the Bayesian joint models are implemented in \texttt{Stan} through the \texttt{rstan} package, using Hamiltonian Monte Carlo (HMC) to draw samples from the joint posterior distribution. The cumulative hazard is approximated by 15-point Gauss--Legendre quadrature.}

Prior distributions are typically chosen to be independent and marginally diffuse \citep{Gelman2013}. Accordingly, normal distributions with mean zero and standard deviation 10 are assigned to the longitudinal and survival coefficients $(\beta_0,\beta_1,\gamma,\alpha)$. The Weibull shape parameter $\nu$ is assigned an inverse-gamma$(0.1,0.1)$ prior. The dispersion parameter $\phi$ is assigned a weakly informative half-Cauchy$(0,5)$ prior \citep{Gelman2006}. The covariance matrix for the random effects, $\Sigma$, is modeled using an inverse-Wishart$(I_2,2)$ distribution, where $I_2$ is a $2\times2$ identity matrix \citep{Schuurman2016}.

\subsection{Dynamic survival predictions}

One of the main advantages of the Bayesian approach in joint modeling is that it provides a natural framework for dynamic predictions, as uncertainty is propagated directly through the posterior distribution. {As additional longitudinal measurements become available for a new subject, they update the conditional distribution of the subject-specific random effects and, consequently, the corresponding survival prediction, while uncertainty in the fitted model parameters is propagated through their posterior distribution. This updating process allows increasingly individualized prognostic information to be obtained from the subject's evolving longitudinal trajectory \citep{Andrinopoulou2021}.}

{The goal of dynamic prediction is to estimate the survival probability for a new subject $l$ who has provided a set of longitudinal measurements up to time $t$, denoted by $\mathcal{Y}_l(t)$. Let $u>t$ denote the prediction horizon, that is, the future time point at which the subject-specific survival probability is evaluated. Based on the previously fitted joint model, this conditional survival probability is defined as:}

{
\begin{equation}\label{eq::dynpred_surv}
\pi_l(u\mid t)
=
\Pr\left(
T_l\geq u
\mid
T_l>t,\mathcal{Y}_l(t),L_l,D_n
\right),
\end{equation}
}

where $D_n$ represents the sample of observations for the $n$ subjects used to fit the joint model.

{This survival probability can be expressed in terms of the random effects and model parameters as:}

{
\begin{equation}\label{eq::dynpred_surv2}
\begin{split}
\pi_l(u\mid t)
={}&
\int\int
\Pr\left(
T_l\geq u
\mid
T_l>t,\bm b_l;\bm\theta
\right)\\
&\quad\times
f\left(
\bm b_l
\mid
T_l>t,\mathcal{Y}_l(t),L_l;\bm\theta
\right)
f(\bm\theta\mid D_n)
\,d\bm b_l\,d\bm\theta\\
={}&
\int\int
\dfrac{
S_l(u\mid\bm b_l;\bm\theta)
}{
S_l(t\mid\bm b_l;\bm\theta)
}
f\left(
\bm b_l
\mid
T_l>t,\mathcal{Y}_l(t),L_l;\bm\theta
\right)
f(\bm\theta\mid D_n)
\,d\bm b_l\,d\bm\theta .
\end{split}
\end{equation}
}

{Here, $S_l(\cdot\mid\bm b_l;\bm\theta)$ represents the subject-specific survival function corresponding to the survival submodel specified in Equation~\eqref{eq:survival_submodel} and detailed in Equation~\eqref{eq::jm_survpart}.}

{An iterative Monte Carlo algorithm can be used to approximate the integral in Equation~\eqref{eq::dynpred_surv2} \citep{Rizopoulos2011}. Let $k=1,\ldots,K$ index the Monte Carlo samples. For each iteration $k$, samples are generated according to the following scheme:}

{
\begin{enumerate}
    \item $\bm\theta^{(k)}
    \sim f(\bm\theta\mid D_n)$, using the posterior MCMC samples obtained during model fitting.

    \item $\bm b_l^{(k)}
    \sim
    f\left(
    \bm b_l
    \mid
    T_l>t,\mathcal{Y}_l(t),L_l;
    \bm\theta^{(k)}
    \right)$.

    \item
    $\displaystyle
    \pi_l^{(k)}(u\mid t)
    =
    \dfrac{
    S_l(u\mid\bm b_l^{(k)};\bm\theta^{(k)})
    }{
    S_l(t\mid\bm b_l^{(k)};\bm\theta^{(k)})
    }.$
\end{enumerate}
}

{The first step uses posterior samples already generated during model fitting, and the third step follows directly from the fitted survival submodel. The second step requires sampling from the conditional posterior distribution of the random effects for the new subject, which has the following nonstandard form:}

{
\begin{equation*}
f\left(
\bm b_l
\mid
T_l>t,\mathcal{Y}_l(t),L_l;
\bm\theta^{(k)}
\right)
\quad\propto
\dfrac{
S_l(t\mid\bm b_l;\bm\theta^{(k)})
}{
S_l(L_l\mid\bm b_l;\bm\theta^{(k)})
}
\prod_{j:t_{lj}\leq t}
f\left(
y_{lj}\mid
\bm b_l;\bm\theta_y^{(k)}
\right)
f\left(
\bm b_l;\theta_b^{(k)}
\right).
\end{equation*}
}

In this case, the Metropolis--Hastings algorithm is employed to draw samples from this posterior distribution of the random effects. This algorithm constructs a Markov chain that iteratively explores the parameter space and whose stationary distribution is the target posterior. This provides a practical way to approximate the distribution of the individual random effects.

{Finally, once the $K$ Monte Carlo samples have been drawn, the survival probability in Equation~\eqref{eq::dynpred_surv} is estimated by the Monte Carlo mean:}

{
\begin{equation*}
\widehat{\pi}_l(u\mid t)
=
\frac{1}{K}
\sum_{k=1}^{K}
\pi_l^{(k)}(u\mid t).
\end{equation*}
}

{The posterior median may alternatively be reported as a point summary, while posterior quantiles can be used to quantify uncertainty in the dynamic survival prediction.}

\section{Data application}\label{Section5}

\subsection{Joint model results}

{We apply the proposed Bayesian joint modeling framework to the COPD study described in Section~\ref{Section2}. The same data have previously been analyzed using longitudinal beta-binomial mixed-effects models \citep{Esteban2020} and the  two-stage strategy described in Section~\ref{Section4} linking the fitted longitudinal trajectories with mortality \citep{Galan-Arcicollar2024}. The use of the same cohort and questionnaire dimensions allows us to compare the longitudinal trend estimates with those obtained when survival is not incorporated into their estimation and to assess how simultaneous estimation affects inference on the association between HRQoL and mortality.}

{As described in Section~\ref{Section2}, HRQoL was assessed using the eight dimensions of the SF-36 and the three dimensions of the SGRQ. Higher SF-36 scores indicate better health status, whereas higher SGRQ scores indicate poorer health status. The analysis is conducted using the rescaled discrete versions of the dimension scores, thereby retaining their original bounded integer structure.}

{The proposed model is fitted separately to each questionnaire dimension, with the longitudinal and survival parameters estimated simultaneously. Table~\ref{tab::JM} presents posterior summaries for the fixed longitudinal slope $\beta_1$, the dispersion parameter $\phi$, the standard deviations of the random intercept and slope, denoted by $\sigma_{b_0}$ and $\sigma_{b_1}$, respectively, and the association parameter $\alpha$.}

\begin{table}[h!]
\resizebox{\textwidth}{!}{
\begin{tabular}{llcccccc}
\hline
& & & \multicolumn{5}{c}{Joint Specification Model COPD Study Results} \\
\cline{4-8}
& & & $\beta_{1}$ & $\phi$ & $\sigma_{b_{0}}$ & $\sigma_{b_{1}}$ & $\alpha$ \\
\hline

SF-36 & PF & (20) & -0.08 (-0.11, -0.06) & 0.02 (0.01, 0.02) & 1.17 (0.98, 1.39) & 0.03 (0.02, 0.04) & -2.98 (-3.78, -2.20) \\
      & RP & (4)  & -0.17 (-0.24, -0.11) & 0.48 (0.40, 0.57) & 3.17 (2.18, 4.32) & 0.09 (0.05, 0.14) & -1.13 (-1.80, -0.48) \\
      & BP & (9)  & -0.03 (-0.07, 0.02) & 0.29 (0.25, 0.33) & 1.51 (1.05, 2.06) & 0.05 (0.03, 0.08) & -1.09 (-2.11, -0.06) \\
      & GH & (20) & -0.05 (-0.07, -0.03) & 0.02 (0.01, 0.02) & 0.81 (0.67, 0.96) & 0.02 (0.02, 0.03) & -1.34 (-2.37, -0.33) \\
      & VT & (20) & -0.05 (-0.08, -0.02) & 0.05 (0.04, 0.06) & 1.23 (1.00, 1.47) & 0.03 (0.03, 0.05) & -2.18 (-3.11, -1.26) \\
      & SF & (8)  & -0.07 (-0.13, -0.01) & 0.19 (0.15, 0.23) & 2.45 (1.72, 3.30) & 0.07 (0.04, 0.11) & -1.63 (-2.51, -0.70) \\
      & RE & (3)  & -0.15 (-0.24, -0.05) & 1.19 (0.96, 1.45) & 3.73 (2.30, 5.74) & 0.13 (0.07, 0.22) & -1.22 (-1.90, -0.53) \\
      & MH & (13) & -0.04 (-0.07, -0.01) & 0.01 (0.01, 0.02) & 1.32 (1.08, 1.59) & 0.04 (0.03, 0.05) & -1.49 (-2.39, -0.57) \\
  &&&&&&&\\

SGRQ & SYMP & (24) & 0.01 (-0.02, 0.04) & 0.06 (0.05, 0.07) & 0.72 (0.57, 0.89) & 0.03 (0.02, 0.04) & 0.80 (-0.19, 1.78) \\
     & IMP & (24) & -0.00 (-0.03, 0.02) & 0.02 (0.01, 0.02) & 1.24 (1.04, 1.46) & 0.03 (0.02, 0.04) & 1.48 (0.67, 2.30) \\
     & ACT & (24) & 0.04 (0.01, 0.07) & 0.03 (0.03, 0.04) & 1.59 (1.33, 1.90) & 0.03 (0.03, 0.04) & 2.12 (1.37, 2.90) \\
\hline
\end{tabular}
}
\caption{Posterior means and 95\% credible intervals of the model parameters for each questionnaire dimension. The maximum score for each dimension is indicated in parentheses.}
\label{tab::JM}
\end{table}

{For the SF-36, all posterior mean slope estimates are negative, indicating a deterioration in health status over time. The 95\% credible intervals exclude zero for all dimensions except \textit{Bodily Pain} (BP). For the SGRQ, the estimated slopes for \textit{Symptoms} (SYMP) and \textit{Impacts} (IMP) are close to zero and their credible intervals contain zero. In contrast, the positive slope for \textit{Activity} (ACT), whose credible interval excludes zero, provides evidence of deterioration in this dimension.}

{Compared with the longitudinal-only beta-binomial estimates used in the two-stage analysis \citep{Galan-Arcicollar2024}, the slope estimates obtained from the simultaneous Bayesian joint model generally have larger absolute values while retaining the same direction. This pattern suggests a more pronounced deterioration than that estimated without incorporating the survival process. Because the application alone cannot establish the source of these differences, they should be interpreted as being consistent with mortality-related informative dropout: patients experiencing earlier events contribute fewer longitudinal observations, and their trajectories may therefore be insufficiently represented when the longitudinal process is fitted independently. The simulation study in Section~\ref{Section6} examines this mechanism directly.}

{The simultaneous model also produces generally larger random-intercept standard deviations and smaller random-slope standard deviations than the longitudinal-only estimates, where the latter are available. Moreover, the previous longitudinal analysis was unable to estimate the random-slope variability for the BP, SF, MH, and IMP dimensions, whereas the simultaneous Bayesian model provides posterior estimates for these components. These differences indicate that the allocation of between-subject heterogeneity changes when the longitudinal and survival processes are estimated jointly, although they should not be interpreted as evidence that either  variance component is systematically under- or overestimated from this application alone.}

{For the survival submodel, the signs of the association estimates are consistent with the scoring directions of the questionnaires. All SF-36 dimensions have negative posterior mean estimates of $\alpha$, indicating that better perceived health is associated with a lower mortality hazard. Conversely, all SGRQ dimensions have positive posterior mean estimates, indicating that poorer perceived health is associated with a higher mortality hazard.}

{The 95\% credible intervals for $\alpha$ exclude zero for all SF-36 dimensions. In comparison, the earlier two-stage analysis found evidence of an association only for \textit{Physical Functioning} (PF) and \textit{Vitality} (VT) \citep{Galan-Arcicollar2024}. For the SGRQ, the credible intervals exclude zero for \textit{Impacts} (IMP) and \textit{Activity} (ACT), whereas the interval for \textit{Symptoms} (SYMP) contains zero. Among the SGRQ dimensions, ACT exhibits the strongest association with mortality. For example, a $0.10$ increase in its underlying probability trajectory $p_i(t)$, corresponding to an increase equal to 10\% of the dimension's maximum possible score, is associated with an approximately 24\% higher mortality hazard, since $\exp(0.10\times2.12)\approx1.24$.}

\subsection{Illustration of dynamic survival predictions}

{To illustrate the applicability of the proposed framework for dynamic prediction, three patients were excluded from model fitting and retained as illustrative cases. The joint model was fitted using the remaining patients, whereas the longitudinal histories of Patients A, B, and C were used to obtain updated survival predictions. These patients were selected to represent different follow-up and longitudinal profiles:}
\begin{itemize}
    \item \textit{Patient A}, who was lost to follow-up and censored after three recorded measurements.
    \item \textit{Patient B}, who was administratively censored, having completed the full follow-up period with all four planned measurements.
    \item \textit{Patient C}, who recorded three measurements before experiencing the event during the follow-up period.
\end{itemize}

\begin{figure}[h!]
    \centering
    \begin{minipage}{0.31\linewidth}
        \includegraphics[width=1\linewidth]{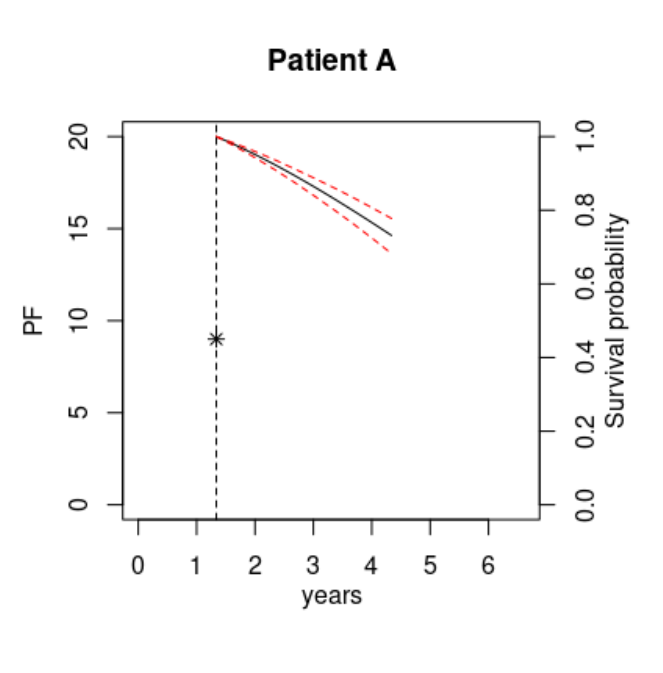}
    \end{minipage}
    \begin{minipage}{0.31\linewidth}
        \includegraphics[width=1\linewidth]{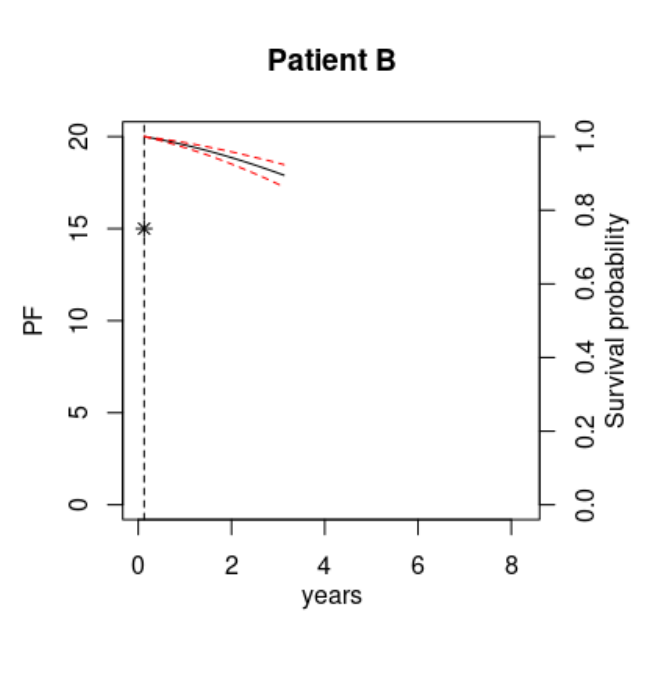}
    \end{minipage}
    \begin{minipage}{0.31\linewidth}
        \includegraphics[width=1\linewidth]{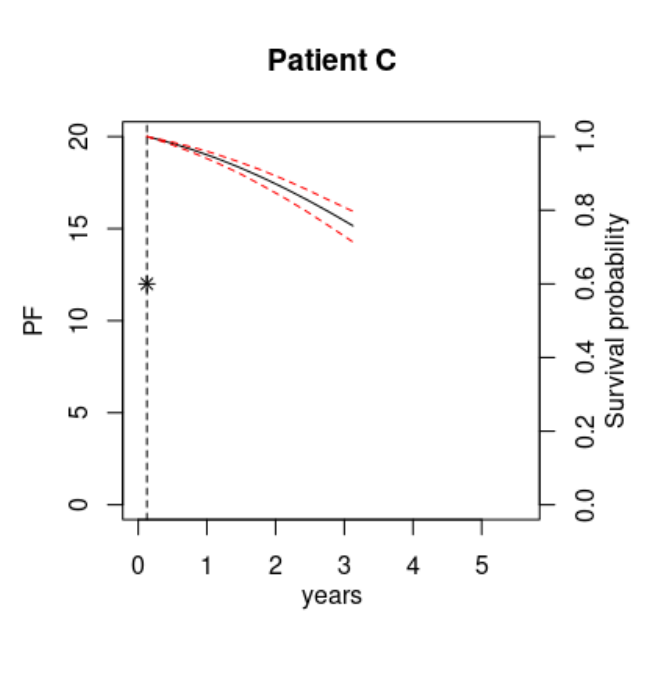}
    \end{minipage}

    \begin{minipage}{0.31\linewidth}
        \includegraphics[width=1\linewidth]{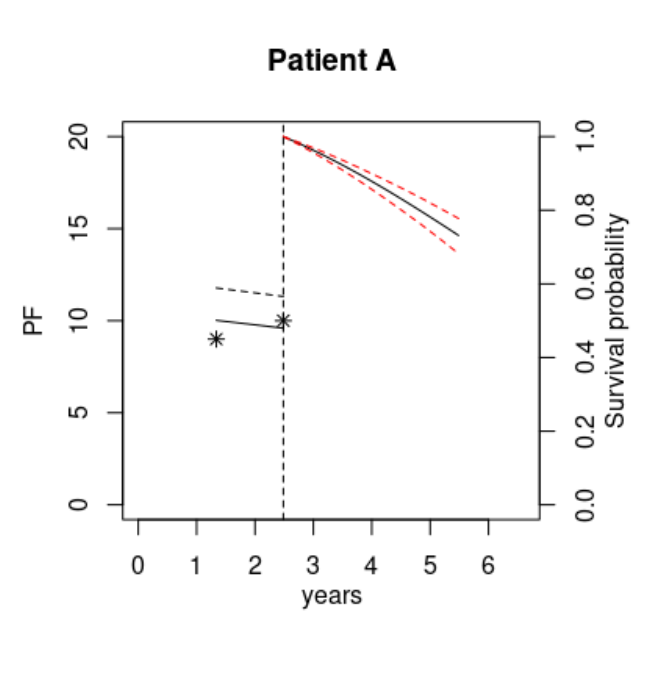}
    \end{minipage}
    \begin{minipage}{0.31\linewidth}
        \includegraphics[width=1\linewidth]{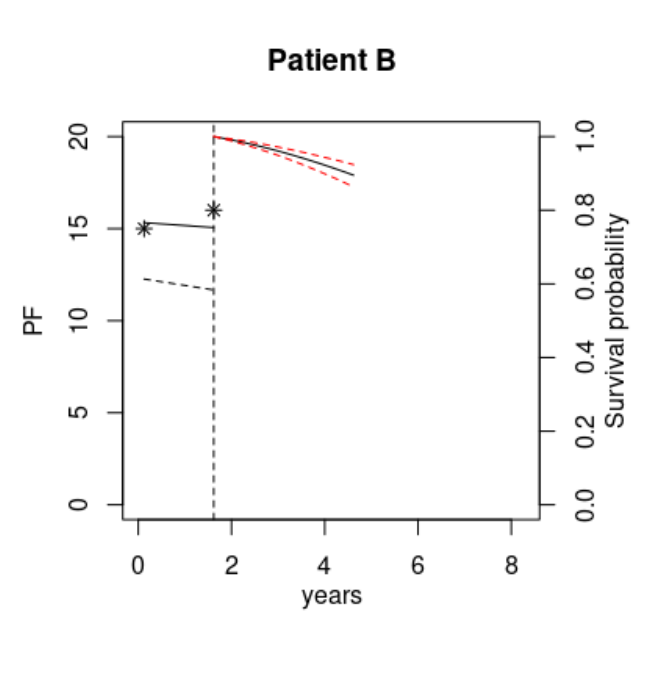}
    \end{minipage}
    \begin{minipage}{0.31\linewidth}
        \includegraphics[width=1\linewidth]{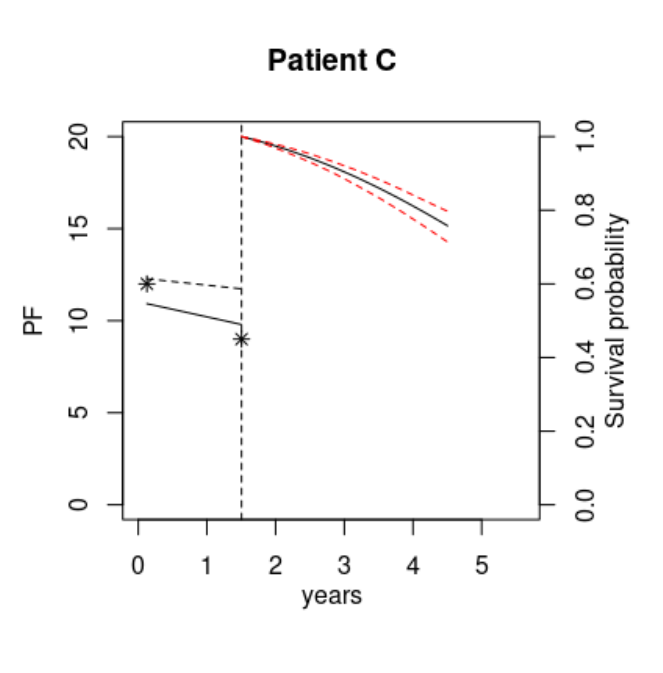}
    \end{minipage}

    \begin{minipage}{0.31\linewidth}
        \includegraphics[width=1\linewidth]{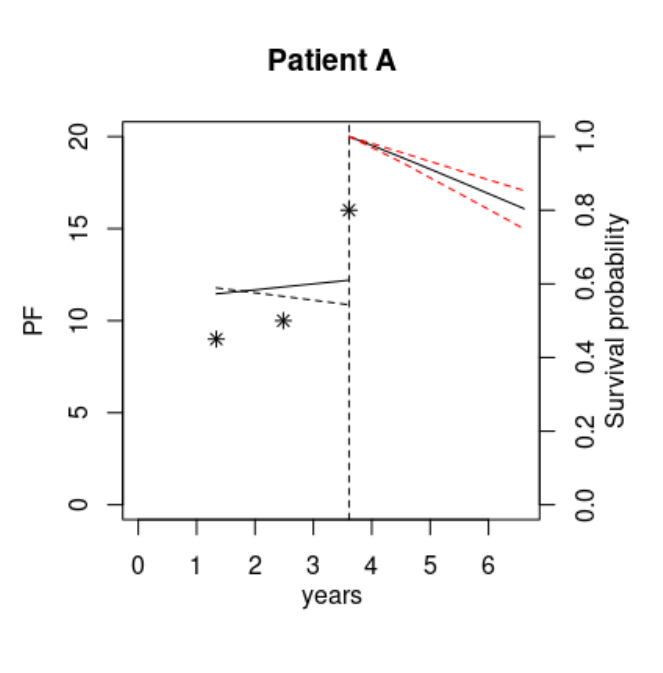}
    \end{minipage}
    \begin{minipage}{0.31\linewidth}
        \includegraphics[width=1\linewidth]{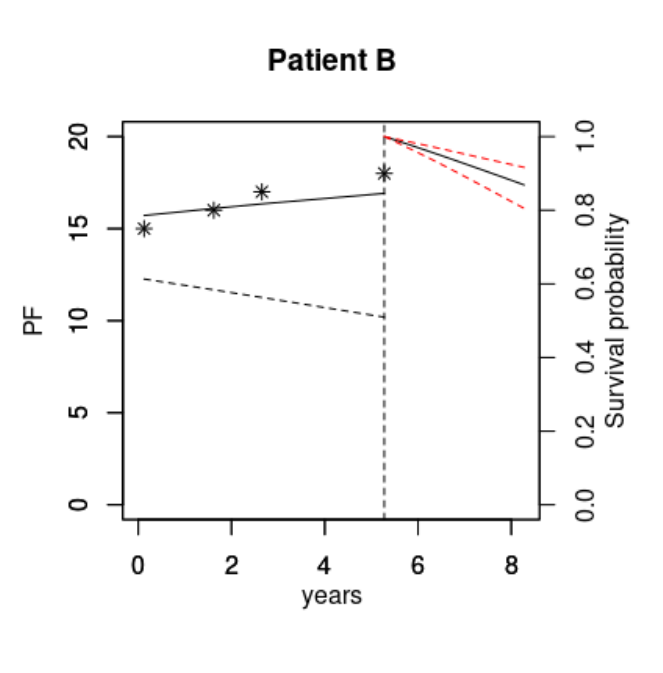}
    \end{minipage}
    \begin{minipage}{0.31\linewidth}
        \includegraphics[width=1\linewidth]{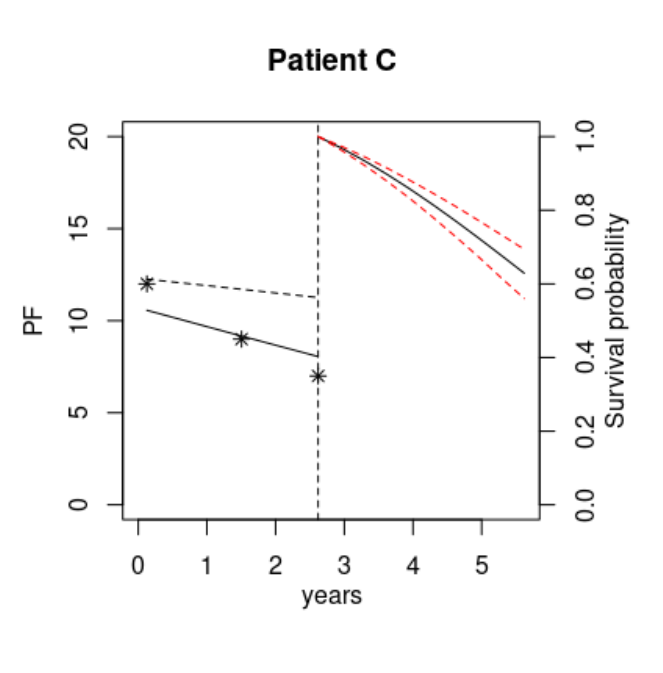}
    \end{minipage}

    \caption{Dynamic survival predictions for three COPD patients based on the \textit{Physical Functioning} (PF) dimension. Columns correspond to Patients A, B, and C, while rows show predictions updated using the longitudinal information available at the first, second, and last measurement, respectively. Time 0 corresponds to the start of the study. On the left side of each panel, the dashed line represents the population trajectory and the solid line the estimated subject-specific trajectory. On the right side, survival probabilities are predicted over the three years following the corresponding landmark time.}
    \label{fig::dynpred}
\end{figure}

{Dynamic survival predictions based on the \textit{Physical Functioning} (PF) dimension of the SF-36 are presented in Figure~\ref{fig::dynpred}. For each patient, predictions are sequentially updated using the longitudinal information available at the first, second, and last measurement. At each landmark time, survival probabilities are predicted over the subsequent three-year period. Only the longitudinal measurements recorded up to the corresponding landmark time and the information that the patient remained event-free until that time are used to obtain each prediction.}

{Patients A and B have positive estimated PF trajectories, whereas Patient C exhibits a negative trajectory. Because higher PF values are associated with a lower mortality hazard, the predicted survival probabilities for Patients A and B are correspondingly higher than those obtained for Patient C. As additional measurements are incorporated, the estimated individual trajectories and survival probabilities are updated to reflect the accumulating patient-specific information.}

{This illustrative comparison shows how differences in longitudinal patterns are reflected in the survival predictions. Even with a limited number of measurements, the estimated subject-specific trajectories reflect positive and negative health evolution, leading to corresponding updates in the predicted survival probabilities.}

\section{Simulation study}\label{Section6}

{In this section, we present the results of a simulation study designed to assess the finite-sample performance of the simultaneous Bayesian joint model (JMBB) in comparison with the two-stage estimation strategy (TSBB). Specifically, we evaluate whether simultaneous estimation reduces bias in the longitudinal slope relative to the longitudinal model fitted independently in the first stage of TSBB and in the association parameter relative to the two-stage plug-in estimator.}

\subsection{Simulation design}

The general settings for the simulation setup are based on the motivating COPD dataset presented in Section \ref{Section2}. {To reproduce the staggered enrollment and delayed entry observed in the COPD study, the entry time $L_i$ is generated from a uniform distribution, $L_i\sim U(0,1.5)$, with the first longitudinal measurement recorded at this time.} The second measurement occurs one year after the first, followed by a third measurement one year later. The fourth measurement is taken three years after the third. {Therefore, the planned measurement times for subject $i$ are $L_i$, $L_i+1$, $L_i+2$, and $L_i+5$, reproducing the unequally spaced measurements and five-year follow-up period of the COPD study.}

Given these measurement times, the longitudinal measurements are simulated based on the longitudinal submodel described in Section \ref{Subsection3.1}.

For survival data, we generate the event times, denoted by $T_{i}^{*}$,
according to the survival submodel described in  Section \ref{Subsection3.2}, by evaluating the inverse of the cumulative hazard, as outlined in the literature \citep{Crowther2013}. Additionally, to account for potential dropouts, a loss-to-follow-up censoring time, denoted by $C_{i}$, is introduced and modeled using a uniform distribution between each subject's first and last recorded measurement. Approximately 10\% of individuals are censored, consistent with the findings from the COPD study. Finally, administrative censoring, denoted by $A_{i}$, is applied to subjects who completed the follow-up period with four recorded longitudinal measurements. The observed survival time is then defined as $T_{i} = \min\{T_{i}^{*},A_{i},C_{i}\}$ with corresponding event indicator $\delta_{i}=1$ if $T_{i}=T_{i}^{*}$ and $\delta_{i}=0$ otherwise. Longitudinal measurements taken after $T_{i}$ are disregarded.

Once the main simulation setup is defined, we divide it into two main scenarios to reflect the two HRQoL questionnaires considered in the COPD study. {The first scenario is based on the SGRQ, for which higher scores indicate poorer health and are associated with a higher mortality risk, whereas the second is based on the SF-36, for which higher scores indicate better health and are associated with a lower mortality risk.} The parameters selected for simulating the datasets are based on the results presented in Section \ref{Section5} and are shown in Table \ref{tab::longitudinalfix_JMBB}. Specifically, Scenario 1 is based on the ACT dimension of the SGRQ, and Scenario 2 on the SF dimension of the SF-36.

\begin{table}[h!]
\centering
  	\begin{tabular}{lcccccccc}
   		\hline
    		&\multicolumn{8}{c}{Model parameters} \\ \cline{2-9}
    		& $\beta_{0}$&$\beta_{1}$&$\sigma_{b_{0}}$&$\sigma_{b_{1}}$ & $m$&$\alpha$ & $\gamma$&$\nu$ \\\hline
    		Scenario 1&-0.20&0.04&1.60& 0.03&24& $>0$& -5.19& 1.75\\
    		&&&&&&&&\\
    		Scenario 2&2.17&-0.07&2.45&0.07&8&$<0$ &-2.77&1.75\\
    		\hline
    	\end{tabular}
	\caption{Parameter values used in the two main simulation scenarios, based on the estimates obtained from the COPD data application.}
	\label{tab::longitudinalfix_JMBB}
\end{table}

{To introduce variability across simulation settings, the association parameter $\alpha$ and the dispersion parameter $\phi$ are allowed to vary across predefined sets of values. This design enables evaluation of the methods under different levels of extra-binomial dispersion and different strengths of association between the longitudinal and survival processes.} Specifically, the values considered for the dispersion parameter are $\phi\in\{0.05,0.5,1\}$, {representing increasing levels of extra-binomial dispersion.} Regarding the association parameter, we evaluate scenarios reflecting null, moderate, and strong associations. For Scenario 1, the considered values are $\alpha\in\{0,2,4\}$, while for Scenario 2 the selected values are $\alpha\in\{0,-1.5,-3\}$.

The strength of the association directly influences the simulated event times. In Scenario 1, stronger positive $\alpha$ values lead to earlier events, reducing the number of longitudinal measurements observed. {In Scenario 2, stronger negative $\alpha$ values reduce the hazard, leading to later event times, more administrative censoring, and a larger number of observed measurements.} Because longitudinal measurements are recorded only while individuals remain event-free, these differences in event timing directly affect the effective follow-up and the observed number of measurements per subject.

For each of the 18 scenarios, we generate 500 independent replications of the dataset, considering a sample size of $n=500$. {To summarize the characteristics of the simulated longitudinal and survival data, Table \ref{tab::simdata_characteristics} presents the average number of events per simulated dataset, the average number of longitudinal measurements per subject, and the percentage of subjects with only one recorded longitudinal measurement.}

    \begin{table}[h!]
	\resizebox{\textwidth}{!}{
		\centering
		\begin{tabular}{llccc c ccc c ccc}
			\hline
			\multicolumn{13}{c}{Simulated data characteristics}\\
			\hline
			$\alpha$ & &
			\multicolumn{3}{c}{Null} &&
			\multicolumn{3}{c}{Moderate} &&
			\multicolumn{3}{c}{Strong}\\
			\cline{3-5}\cline{7-9}\cline{11-13}

    	& & Events & Meas & {$\%\,n_i=1$} &&
			Events & Meas & {$\%\,n_i=1$} &&
			Events & Meas & {$\%\,n_i=1$}\\
			\hline

    	Scenario	1 &	 &49.25 &
        3.61& 30.39 && 129.05 &
        3.39 & 40.51 && 256.96 &
        2.92 & 76.49   \\
    	Scenario	2 &	 &325.80 &
        2.73 & 82.61  && 154.71 &
        3.31 & 43.97  && 68.13 &
        3.57 & 31.93 \\

    		\hline
    	\end{tabular}
    }
    	\caption{Characteristics of the simulated datasets under null, moderate, and strong associations. Events denotes the average number of observed events per simulated dataset, Meas the average number of longitudinal measurements per subject, and $\%\,n_i=1$ the average percentage of subjects with only one recorded longitudinal measurement.}
    	\label{tab::simdata_characteristics}
    \end{table}

    {The two-stage strategy is implemented using the \texttt{BBmm} function from the \texttt{PROreg} R package \citep{PROreg} for the longitudinal submodel and the \texttt{coxph} function from the \texttt{survival} R package \citep{survival-book} for the survival submodel.} To implement the joint specification approach, we developed Stan code and used the \texttt{rstan} R package \citep{rstan2024}, setting the MCMC configuration to three chains of 2000 iterations each, with a warm-up of 1000. Replicates in which $\alpha$ showed $\hat{R}>1.05$ or an effective sample size below 600 were discarded and regenerated until 500 convergent replicates were obtained for each scenario. The code implementing our proposed method is available on GitHub \url{https://github.com/cgalanarcicollar/JMBB} .

 {Both approaches provide estimates of all longitudinal submodel parameters and of the association parameter $\alpha$. The simultaneous Bayesian joint model additionally estimates the Weibull baseline hazard parameters $\gamma$ and $\nu$, whereas the two-stage approach estimates $\alpha$ through Cox partial likelihood and therefore does not require a parametric specification of the baseline hazard.}

\subsection{Results}

{For the longitudinal submodel, we focus on the fixed slope $\beta_1$, as characterizing temporal change is a central objective in longitudinal studies and allows us to evaluate the effect of mortality-related dropout. Results for $\beta_1$ under the two main scenarios are reported in Tables \ref{tab::S1_beta1_500} and \ref{tab::S2_beta1_500}. For the survival submodel, we focus on the association parameter $\alpha$, which quantifies the relationship between the longitudinal trajectory and mortality risk. Corresponding results for $\alpha$ are presented in Tables \ref{tab::S1_alpha_500} and \ref{tab::S2_alpha_500}.}

{All tables report relative bias (RB), empirical standard deviation (ESD), average estimated standard deviation (ASD), and empirical coverage probability (CP) of the nominal 95\% intervals. Relative bias is defined as $\operatorname{RB}(\widehat{\theta})=(\overline{\widehat{\theta}}-\theta)/\theta$, where $\theta$ is the true parameter value and $\overline{\widehat{\theta}}$ is the average estimate across the simulation replications. Because relative bias is undefined when $\alpha=0$, signed bias is reported for the null-association settings. Coverage is evaluated using confidence intervals for TSBB and posterior credible intervals for JMBB. Additionally, Figure \ref{fig::boxplot_beta1} presents boxplots of the relative bias values for $\beta_1$, while Figure \ref{fig::boxplot_alpha} presents relative bias values for $\alpha$ under the non-null association settings and signed bias values when $\alpha=0$.}

\subsubsection{Longitudinal results}

    \begin{table}[h!]
   	    \resizebox{\textwidth}{!}{
    	\centering
    	\begin{tabular}{p{0.2cm}rp{1.2cm}rrrrrrrrrrrrrr}
    		\hline
    		&    &   &  \multicolumn{14}{c}{Results for $\beta_1$ parameter in Scenario 1}                    \\ \hline
    		&&& \multicolumn{5}{c}{$\alpha= 0$}&\multicolumn{5}{c}{$\alpha= 2$}&\multicolumn{4}{c}{$\alpha= 4$}\\ \cline{4-7}\cline{9-12}\cline{14-17}

    		$\phi$ & & &RB   & ESD    & ASD & CP   &&  RB   & ESD    & ASD & CP  && RB   & ESD    & ASD& CP    \\

    		\hline

    		0.05& &JMBB&0.00 & 0.012 & 0.014 & 97.6 && 0.06 & 0.014 & 0.015 & 97.6&& 0.31 & 0.018 & 0.020 & 92.6 \\

    		&& TSBB& -0.04 & 0.011 & 0.009 & 88.2  && -0.24 & 0.012 & 0.009 & 76.6 && -0.95 & 0.016 & 0.011 & 15.8 \\
    		&&&&&&&&&&&&&&&&\\

    		0.5&& JMBB& 0.00 & 0.020 & 0.022 & 95.6  && 0.12 & 0.021 & 0.024 & 96.8&&  0.66 & 0.029 & 0.033 & 90.4\\
    		&& TSBB&  -0.08 & 0.018 & 0.017 & 93.0 && -0.44 & 0.018 & 0.018 & 81.0 && -1.70 & 0.023 & 0.021 & 12.0 \\
    		&&&&&&&&&&&&&&&&\\

    		1&&JMBB &0.04 & 0.023 & 0.025 & 97.0  &&  0.21 & 0.026 & 0.028 & 96.2 && 0.75 & 0.035 & 0.038 & 89.6 \\
    		&& TSBB&  -0.04 & 0.020 & 0.020 & 95.4 &&-0.41 & 0.022 & 0.022 & 87.8  && -1.76 & 0.027 & 0.026 & 22.0  \\
    		\hline
    	\end{tabular}
    }
        \caption{
    	{
        Simulation results for the fixed slope parameter $\beta_1$ obtained with JMBB and TSBB under Scenario 1, considering $\alpha\in\{0,2,4\}$ and $\phi\in\{0.05,0.5,1\}$. Relative bias (RB), empirical standard deviation (ESD), average estimated standard deviation (ASD), and empirical coverage probability (CP) of the 95\% intervals are reported.}}
        \label{tab::S1_beta1_500}
    \end{table}

    \begin{table}[h!]
        \resizebox{\textwidth}{!}{
    	\centering
    	\begin{tabular}{p{0.2cm}rp{1.2cm}rrrrrrrrrrrrrr}
    		\hline
    		&   &    &  \multicolumn{14}{c}{Results for $\beta_1$ parameter in Scenario 2}                    \\ \hline
    		&&& \multicolumn{5}{c}{$\alpha= 0$}&\multicolumn{5}{c}{$\alpha= -1.5$}&\multicolumn{4}{c}{$\alpha= -3$}\\  \cline{4-7}\cline{9-12}\cline{14-17}

    		$\phi$ && & RB   & ESD    & ASD &CP  &&  RB   & ESD    & ASD & CP && RB   & ESD    & ASD & CP   \\
    		\hline

    		0.05& &JMBB&   -0.26 & 0.040 & 0.042 & 94.8 && -0.15 & 0.028 & 0.030 & 95.6 &&-0.15 & 0.026 & 0.026 & 95.0  \\

    		&& TSBB& -0.17 & 0.034 & 0.022 & 75.4 &&    -0.25 & 0.024 & 0.017 & 74.0 && -0.20 & 0.021 & 0.015 & 76.6\\
    		&&&&&&&&&&&&&&&&\\

    		0.5& &JMBB&-0.44 & 0.055 & 0.058 & 91.4 && -0.26 & 0.038 & 0.040 & 93.2 &&-0.23 & 0.033 & 0.036 & 93.8\\
    		&& TSBB& -0.46 & 0.048 & 0.034 & 75.8   &&  -0.51 & 0.033 & 0.026 & 69.4  && -0.42 & 0.029 & 0.024 & 74.4 \\
    		&&&&&&&&&&&&&&&&\\

    		1&&JMBB & -0.65 & 0.064 & 0.067 & 90.2 &&-0.37 & 0.044 & 0.047 & 92.4 && -0.34 & 0.040 & 0.041 & 93.0\\
    		&& TSBB& -0.76 & 0.057 & 0.041 & 67.8 && -0.70 & 0.038 & 0.032 & 63.0 && -0.60 & 0.034 & 0.029 & 66.4 \\
    		\hline
    	\end{tabular}
    }

   \caption{
   	{
    Simulation results for the fixed slope parameter $\beta_1$ obtained with JMBB and TSBB under Scenario 2, considering $\alpha\in\{0,-1.5,-3\}$ and $\phi\in\{0.05,0.5,1\}$. Relative bias (RB), empirical standard deviation (ESD), average estimated standard deviation (ASD), and empirical coverage probability (CP) of the 95\% intervals are reported.}}
    \label{tab::S2_beta1_500}
    \end{table}

    \begin{figure}[h!]
    	\centering
    	\includegraphics[width=1\linewidth]{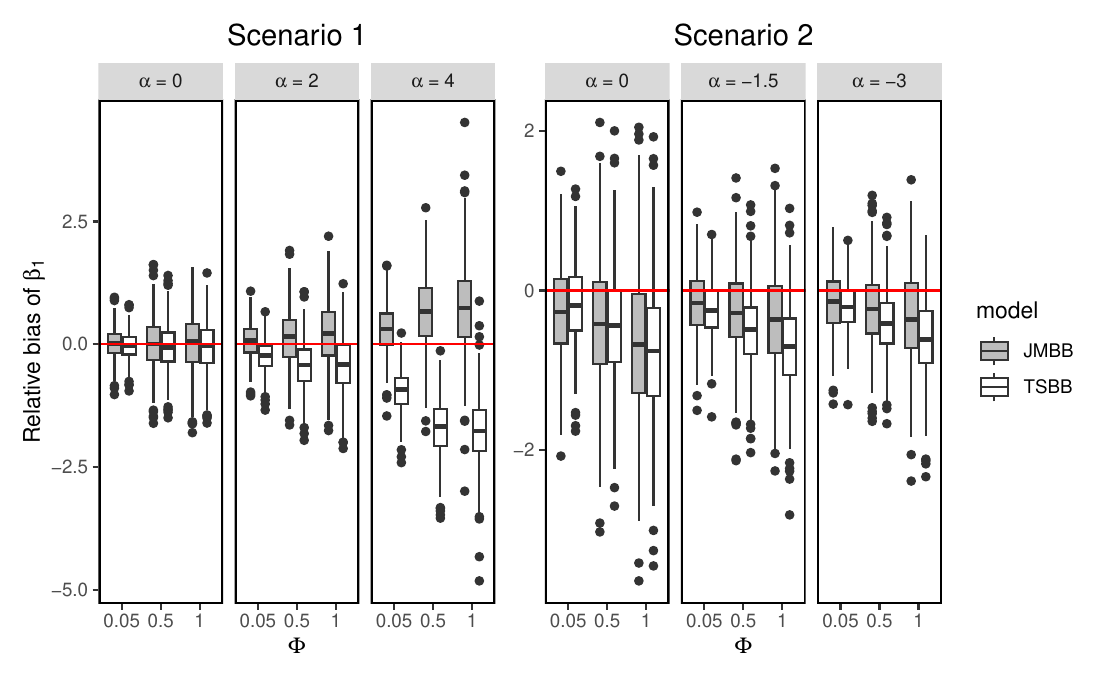}
    	\caption{
    	{
        Boxplots of the relative bias values for $\beta_1$ under Scenario 1 (left) and Scenario 2 (right).}}
        \label{fig::boxplot_beta1}
    \end{figure}

{In Scenario 1 (Table \ref{tab::S1_beta1_500}), both approaches show little bias when there is no association between the longitudinal and survival processes. As $\alpha$ increases, earlier events lead to fewer observed longitudinal measurements and make estimation of the longitudinal trend more difficult. Increasing $\phi$ introduces additional extra-binomial dispersion and further increases the magnitude of the bias.}

{The two approaches behave differently as the association becomes stronger. TSBB increasingly underestimates the positive longitudinal slope because its longitudinal submodel is fitted independently and therefore does not use the survival information associated with mortality-related dropout. JMBB remains closer to the true value, although it shows some upward bias under the strongest association, particularly for larger values of $\phi$. This pattern is also reflected in the interval estimates: JMBB coverage ranges from 89.6\% to 97.6\% and remains near the nominal level except under the most severe association and dispersion settings, whereas TSBB coverage deteriorates substantially as the association increases, reaching values between 12.0\% and 22.0\% when $\alpha=4$.}

{In Scenario 2 (Table \ref{tab::S2_beta1_500}), both approaches tend to estimate slopes closer to zero than the true value $\beta_1=-0.07$, thereby underestimating the deterioration in health status. As the negative association becomes stronger, events occur later and more longitudinal measurements are observed, generally reducing the bias. JMBB produces lower relative bias than TSBB in most settings, although this advantage is not observed in every null-association setting. More importantly, JMBB achieves coverage probabilities between 90.2\% and 95.6\%, whereas those of TSBB range from 63.0\% to 76.6\%. The lower TSBB coverage is consistent with its greater bias and, particularly in Scenario 2, with its average estimated standard deviations being smaller than the empirical variability of the estimates.}

{The distributions displayed in Figure \ref{fig::boxplot_beta1} support these findings. In Scenario 1, the TSBB relative bias becomes increasingly negative as the association strengthens, while JMBB remains closer to zero despite some positive bias in the most demanding settings. In Scenario 2, the relative bias is predominantly negative for both approaches, consistently indicating estimation of a less pronounced deterioration than the one used to generate the data. Nevertheless, TSBB consistently exhibits a larger bias than JMBB, showing that simultaneous estimation reduces, although does not completely eliminate, the bias in the longitudinal slope.}

{These findings are consistent with the pattern observed in the COPD application, where TSBB estimated a less pronounced deterioration than JMBB. The simulation study therefore suggests that fitting the longitudinal process independently may contribute to underestimation of temporal health deterioration when mortality-related dropout is informative.}

\subsubsection{Survival results}

    \begin{table}[h!]
   	    \resizebox{\textwidth}{!}{
    	\centering
    	\begin{tabular}{p{0.2cm}rp{1.2cm}rrrrrrrrrrrrrr}
    		\hline
    		&    &   &  \multicolumn{14}{c}{Results for $\alpha$ parameter in Scenario 1}                    \\ \hline
    		&&& \multicolumn{5}{c}{$\alpha= 0$}&\multicolumn{5}{c}{$\alpha= 2$}&\multicolumn{4}{c}{$\alpha= 4$}\\ \cline{4-7}\cline{9-12}\cline{14-17}

    		$\phi$ & & & $Bias^{*}$   & ESD    & ASD & CP   &&  RB   & ESD    & ASD & CP  && RB   & ESD    & ASD& CP    \\

    		\hline

    		0.05& &JMBB&-0.05 & 0.595 & 0.599 & 94.2 && -0.00 & 0.401 & 0.392 & 95.6&&  0.03 & 0.344 & 0.341 & 93.6 \\

    		&& TSBB&-0.22 & 0.490 & 0.539  & 95.2  && -0.25 & 0.329 & 0.348 & 69.2 && -0.15 & 0.301 & 0.286 & 41.2 \\
    		&&&&&&&&&&&&&&&&\\

    		0.5&& JMBB& -0.02 & 0.703 & 0.689 & 93.8  && 0.04 & 0.469 & 0.459 & 94.2 &&  0.07 & 0.509 & 0.475 & 90.2 \\
    		&& TSBB&  -0.19 & 0.487 & 0.548 & 96.6 && -0.33 & 0.336 & 0.350 & 52.2 && -0.24 & 0.381 & 0.285 & 16.4 \\
    		&&&&&&&&&&&&&&&&\\

    		1&&JMBB &-0.05 & 0.741 & 0.726 & 94.6  &&  0.03 & 0.481 & 0.487 & 95.8&&   0.03 & 0.499 & 0.513 & 95.4 \\
    		&& TSBB&   -0.21 & 0.465 & 0.529 & 96.8 &&  -0.42 & 0.301 & 0.336 & 28.0   &&  -0.35 & 0.354 & 0.268 & 2.2 \\
    		\hline
    	\end{tabular}
    }
        \caption{
        {
        Simulation results for the association parameter $\alpha$ obtained with JMBB and TSBB under Scenario 1, considering $\alpha\in\{0,2,4\}$ and $\phi\in\{0.05,0.5,1\}$. Relative bias (RB), empirical standard deviation (ESD), average estimated standard deviation (ASD), and empirical coverage probability (CP) of the 95\% intervals are reported. $^{*}$Signed bias is reported because relative bias is undefined when $\alpha=0$.}}
    	\label{tab::S1_alpha_500}
    \end{table}

    \begin{table}[h!]
        \resizebox{\textwidth}{!}{
    	\centering
    	\begin{tabular}{p{0.2cm}rp{1.2cm}rrrrrrrrrrrrrr}
    		\hline
    		&   &    &  \multicolumn{14}{c}{Results for $\alpha$ parameter in Scenario 2}                    \\ \hline
    		&&& \multicolumn{5}{c}{$\alpha= 0$}&\multicolumn{5}{c}{$\alpha= -1.5$}&\multicolumn{4}{c}{$\alpha= -3$}\\  \cline{4-7}\cline{9-12}\cline{14-17}

    		$\phi$ && & $Bias^{*}$   & ESD    & ASD &CP  &&  RB   & ESD    & ASD & CP && RB   & ESD    & ASD & CP   \\
    		\hline

    		0.05& &JMBB&    0.03 & 0.291 & 0.269 & 93.8  &&-0.02 & 0.339 & 0.332 & 94.2 && 0.01 & 0.480 & 0.465 & 95.2  \\

    		&& TSBB& 0.47 & 0.233 & 0.230 & 44.8 &&   -0.55 & 0.283 & 0.294 & 17.8 && -0.37 & 0.384 & 0.399 & 19.8\\
    		&&&&&&&&&&&&&&&&\\

    		0.5& &JMBB& 0.02 & 0.307 & 0.298 & 94.4 &&-0.01 & 0.377 & 0.366 & 93.6  &&	0.02 & 0.565 & 0.521 & 92.4\\
    		&& TSBB& 0.42 & 0.210 & 0.217 & 50.6  &&  -0.63 & 0.266 & 0.284 & 6.6   && -0.46 & 0.371 & 0.388 & 4.4  \\
    		&&&&&&&&&&&&&&&&\\

    		1&&JMBB & 0.02 & 0.312 & 0.308 & 95.2 &&-0.01 & 0.375 & 0.379 & 94.8 && -0.00 & 0.579 & 0.544 & 94.6\\
    		&& TSBB&  0.38 & 0.202 & 0.204 & 56.0 &&  -0.67 & 0.246 & 0.271 & 2.2 && -0.52 & 0.358 & 0.373 & 0.8 \\
    		\hline
    	\end{tabular}
    }

   \caption{
   {
    Simulation results for the association parameter $\alpha$ obtained with JMBB and TSBB under Scenario 2, considering $\alpha\in\{0,-1.5,-3\}$ and $\phi\in\{0.05,0.5,1\}$. Relative bias (RB), empirical standard deviation (ESD), average estimated standard deviation (ASD), and empirical coverage probability (CP) of the 95\% intervals are reported. $^{*}$Signed bias is reported because relative bias is undefined when $\alpha=0$.}}
   	\label{tab::S2_alpha_500}
    \end{table}

    \begin{figure}[h!]
    	\centering
    	\includegraphics[width=1\linewidth]{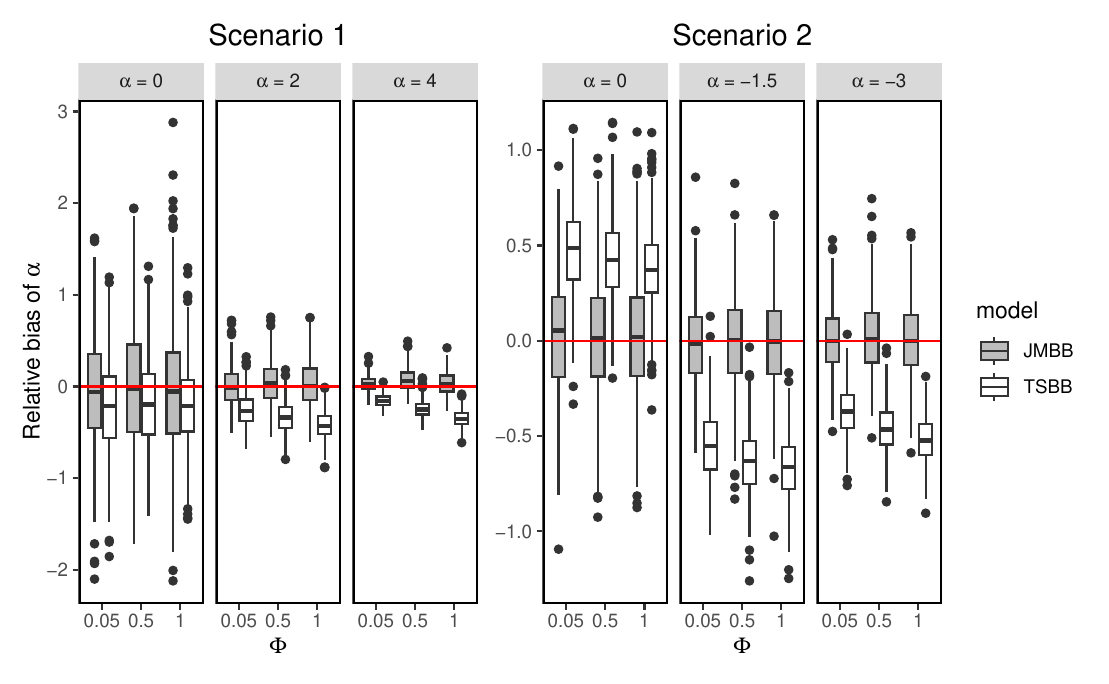}
    	\caption{
        {
        Boxplots of the relative bias values for $\alpha$ under the non-null association settings and of the signed bias values when $\alpha=0$. Scenario 1 and Scenario 2 are displayed from left to right, respectively.}}
    	\label{fig::boxplot_alpha}
    \end{figure}

    {Tables \ref{tab::S1_alpha_500} and \ref{tab::S2_alpha_500} show that JMBB produces practically unbiased estimates of the association parameter regardless of variations in the dispersion parameter or the strength of the association. Under the non-null settings, the relative bias of JMBB does not exceed 0.07 in Scenario 1 or 0.02 in Scenario 2. Its signed bias also remains close to zero when no association is present.}

    {By contrast, TSBB tends to underestimate the magnitude of the association parameter under the non-null settings. In Scenario 1, its relative bias ranges from $-0.16$ to $-0.42$, indicating underestimation of the positive association. In Scenario 2, relative bias ranges from $-0.37$ to $-0.67$. Because the true associations are negative in this scenario, these negative relative biases correspond to estimates that are less negative than the true values and therefore underestimate the magnitude of the association. Figure \ref{fig::boxplot_alpha} clearly displays this displacement of the TSBB estimates relative to JMBB.}

    {The coverage results reinforce these findings. For JMBB, coverage remains close to the nominal level, ranging from 90.2\% to 95.8\% in Scenario 1 and from 92.4\% to 95.2\% in Scenario 2. For TSBB, coverage is close to the nominal level only under the null-association settings of Scenario 1 and deteriorates markedly when an association is present. In Scenario 2, TSBB coverage is low even when $\alpha=0$ and decreases to between 0.6\% and 19.6\% under the non-null settings.}

    {We also observe in both tables that TSBB generally yields lower ESD and ASD than JMBB. The lower ASD is consistent with the plug-in nature of TSBB, which treats the estimated longitudinal trajectory as known and does not propagate its estimation uncertainty into the survival submodel. However, its lower empirical variability is accompanied by substantial attenuation bias. The ESD and ASD values for TSBB are broadly comparable in many settings, although ASD is smaller than ESD in some strong-association settings. Therefore, the poor coverage of TSBB is largely driven by bias, with underestimation of uncertainty also contributing in some scenarios. Simultaneous estimation provides a more reliable balance between bias and uncertainty for inference on the association parameter.}

    {These findings are consistent with the COPD application. In that analysis, TSBB provided evidence of an association for only two of the eight SF-36 dimensions, whereas the 95\% credible intervals obtained with JMBB excluded zero for all eight dimensions. The underestimation of the association observed for TSBB in the simulation study provides a plausible explanation for why some weak-to-moderate associations may not have been identified by the two-stage analysis.}

\section{Conclusions and further research}\label{Section7}

Patient-centered care has increasingly emphasized the role of patients as active participants in their own health management \citep{Wolff2025}, and PROs are highly recommended tools for clinical assessment. In this context, evaluating the temporal evolution of patients' health-related quality of life is important in its own right, while understanding how these trajectories relate to mortality risk provides valuable prognostic information.

Joint models are powerful tools for assessing longitudinal outcomes and their association with mortality. However, the existing literature often overlooks the distributional characteristics of PROs when jointly analyzing longitudinal PRO and survival data. Since PROs are collected through item-based questionnaires and constructed as (weighted) sums of item responses, they are discrete and bounded and frequently exhibit overdispersion. The beta-binomial distribution has been proposed as an appropriate model for capturing these features \citep{Arostegui2007}, and a recent study incorporated it into the joint modeling framework using a two-stage approach \citep{Galan-Arcicollar2024}. The main limitation of this strategy is that the longitudinal process is estimated without using survival information, after which its fitted values are treated as known in the survival submodel. Consequently, the two-stage approach may produce biased longitudinal trend estimates when dropout is related to mortality and attenuated estimates of the association between the two processes \citep{Ye2008}.

To account for the distributional characteristics of PROs and the limitations of two-stage estimation, we adopt a simultaneous Bayesian joint model (JMBB) incorporating a beta-binomial longitudinal submodel. JMBB estimates the parameters of both submodels simultaneously, allowing survival information to contribute to estimation of the longitudinal trajectory and propagating uncertainty across the two processes. Moreover, posterior sampling facilitates subject-specific dynamic predictions, allowing a patient's survival probabilities to be updated as new longitudinal PRO measurements become available.

In this study, we applied the simultaneous Bayesian joint modeling framework to analyze the relationship between HRQoL and mortality in a COPD study conducted by researchers at Galdakao Hospital, Spain. Using two widely used questionnaires, the SF-36 and the SGRQ, we examined the relationships between the different questionnaire dimensions and patients' risk of death. Our findings highlight that the dimensions related to physical functioning and activity (PF and ACT), together with vitality (VT), show the strongest associations with mortality. As an illustration of the model's potential, we also provided dynamic survival predictions for three patients with different profiles, updating predictions as additional longitudinal measurements were incorporated.

Building on the COPD analysis, we conducted a simulation study to evaluate the consequences of simultaneous estimation relative to the two-stage strategy. For the association parameter $\alpha$, JMBB produced practically unbiased estimates and coverage probabilities close to the nominal level across the considered scenarios, whereas TSBB attenuated both positive and negative associations toward zero and exhibited poor coverage under the non-null association settings. For the longitudinal slope $\beta_1$, JMBB generally reduced bias, particularly under non-null associations, and provided substantially better coverage than the independently fitted longitudinal component of TSBB, although coverage fell slightly below the nominal level in the most severe dropout settings. Nevertheless, some residual slope bias remained in settings with limited longitudinal information, early events, and high extra-binomial dispersion.

Several limitations should be acknowledged. First, the limited number of longitudinal measurements per patient (at most four) restricts the precision of the estimated individual trajectories, which may affect both model fit and dynamic predictions. Second, our dynamic predictions were presented as illustrative examples rather than as a formal assessment of predictive performance. Third, the current model considers only a single longitudinal outcome, which may not fully capture the multidimensional nature of HRQoL. Fourth, the simulation study compares a frequentist two-stage strategy using a semiparametric Cox model with a simultaneous Bayesian model assuming a Weibull baseline hazard, with data generated under the latter specification. Although the Cox model remains compatible with the generated proportional-hazards structure, the two approaches differ in their treatment of the baseline hazard and in their inferential paradigms. Consequently, the observed gains cannot be attributed exclusively to simultaneous estimation, and robustness to baseline-hazard misspecification remains to be assessed.

Future work could extend the joint modeling framework to incorporate multiple longitudinal outcomes simultaneously, thereby providing a more comprehensive assessment of patients' health trajectories. Such extensions would introduce additional computational, dependence-modeling, and interpretability challenges that warrant further investigation. The framework could also be extended to include clinically relevant baseline and time-varying covariates and nonlinear longitudinal trajectories.

A formal evaluation of predictive performance is also required before clinical implementation. This should include dynamic measures such as time-dependent AUC and Brier scores, together with internal or external validation strategies to assess the calibration and generalizability of the dynamic predictions. Further simulation studies should examine robustness to misspecification of the baseline hazard and compare simultaneous and two-stage estimation under more closely aligned survival specifications.

In conclusion, simultaneous Bayesian joint modeling provides a coherent framework for analyzing the relationship between longitudinal PROs and survival. By incorporating a beta-binomial longitudinal submodel and estimating both processes simultaneously, it accounts for the discrete, bounded, and overdispersed nature of PROs while reducing the biases observed under independent longitudinal and two-stage estimation. In our simulations, JMBB produced practically unbiased association estimates and generally improved estimation of the longitudinal trend, while in the COPD application it provided evidence of associations that were not identified by the two-stage analysis.

\subsubsection*{Acknowledgments}
This research is supported by AEI and EJ-GV under Grants PID2020-115882RB-I00 / AEI / 10.13039 /  501100011033 with acronym “S3M1P4R” and PID2023-153222OB-I00 granted by MCIU /AEI / 10.13039/ 501100011033 /FEDER, UE with acronym ``SPHERES'', Departamento de Educaci\'on, Pol\'itica Ling\"u\'istica y Cultura del Gobierno Vasco [IT1456-22], by the Basque Government through the BERC 2022-2025 program and by the Ministry of Science and Innovation: BCAM Severo Ochoa accreditation CEX2021-001142-S / MICIN / AEI / 10.13039 /  501100011033, by the U.K. Medical Research Council (MRC) grant MC\_UU\_00002/5, and by MRC Unit Theme number MC\_UU\_00040/02 (Precision Medicine). We also gratefully acknowledge Dr. Cristobal Esteban for providing the COPD data and Inmaculada Arostegui for the scientific discussion.

\subsubsection*{Conflict of interest}

The authors declare no potential conflicts of interest.

\bibliographystyle{apalike}
\bibliography{paperreferences}

\phantom{aaaa}

\end{document}